\documentclass[epj]{svjour}
\usepackage{graphics,epsfig}
\begin{document}
\title{An introduction to the Generalized Parton Distributions}
\author{Michel Gar\c con
}                     
\institute{DAPNIA/SPhN, CEA-Saclay, 91191 Gif-sur-Yvette, France}
\date{Received: date / Revised version: date}
%
\abstract{
The concepts of Generalized Parton Distributions (GPD)
are reviewed in an introductory and phenomenological fashion.
These distributions provide a rich and unifying picture of the nucleon
structure. Their physical meaning is discussed.
The GPD are in principle measurable through exclusive deeply virtual
production of photons (DVCS) or of mesons (DVMP). Experiments are starting to
test the validity of these concepts. First results are discussed and new
experimental projects presented, with an emphasis on this program at
Jefferson Lab.
\PACS{
      {12.38.-t}{Quantum chromodynamics} \and   
      {13.60.-r}{Photon and charged-lepton interactions with hadrons} 
     } 
} 
\maketitle
{\sl
	\underline{In Memoriam} :
	Ren\'e Beurtey died a few days before this conference began.
	His work has had a determining influence on the technical and physical 
	developments in our field. His ideas were innovative and he shared them
	generously, not being concerned with ``publishing first".
	He developed
	the first adiabatic transitions in a polarized deuteron source,
	promoted original technics for the study of nucleon-nucleon
	scattering as a continuous function of energy (cluster jet inside
	a synchrotron ring and the ``Beurtey-Saudinos" wheel), and triggered the
	``Tatischeff experiments" on narrow dibaryons search. 
	In addition, he participated
	in the accelerator research that made SATURNE the first synchrotron
	to successfully cross proton depolarizing resonances. 
	Beyond his professional talents, Ren\'e had an expansive and
	enjoyable personality that made him a true pleasure to work
	with or around, or to consult. His lively wit and sense of humor
	were a stimulation for many of us.
}	
\section{Foreword}
This presentation is intended to provide a pedagogical introduction to the 
subject of Generalized Parton Distributions and associated phenomenology.
It cannot be exhaustive on the subject, and the reader should consult
reviews or papers such as Refs.~\cite{Vdh99,Fil01,Goe01,Rad01,Bel02},
and references therein, 
for a more rigourous and complete account of the subject. 
At this conference, M. Vanderhaeghen gave another talk on the subject,
with more emphasis on theory and its latest developments~\cite{Vdh02}.

\section{Basic concepts}
\label{sec:concept}
\begin{figure}
\vspace{1mm}
\centerline{\epsfig{file=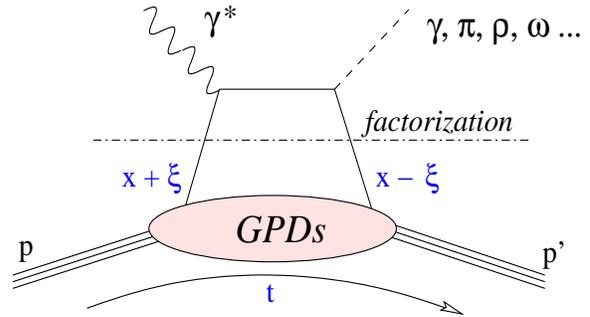,height=42mm,width=77mm}}
\vspace{2mm}
\caption{Representation of the deeply virtual 
	 exclusive processes DVCS and DVMP, also referred
	 to as the ``handbag" diagram.
	}
\label{fig:handbag}
\end{figure}

In the last few years, significant theoretical advances were made that
allow, using Quantum Chromodynamics (QCD) and within some conditions, 
to factorize the amplitude for exclusive 
electroproduction processes
into a hard scattering at the parton level
and non-perturbative {\sl Generalized Parton Distributions} (GPD) - 
see Fig.~\ref{fig:handbag}. 
From the theoretical point of view, as a particular case
of factorization theorems, the introduction of these distributions builds
a bridge between fundamental QCD, phenomenology and experimental observables.
The processes under consideration are 
the deeply virtual Compton scattering (DVCS: $ep\to ep\gamma$) and 
the deeply virtual
meson production (DVMP: $ep\to ep\ \rho,\omega,\pi...$).
We understand by ``deeply virtual" a kinematical regime where
the virtuality $Q^2$ of the exchanged photon $\gamma^*$ is large 
 and where the
available energy in the $\gamma^*p$ system is above the resonance region,
typically larger than 2 GeV.

The GPDs contain the hitherto most complete description of the internal 
quark-gluon structure of hadrons.
For each of the quark flavours (we will not consider the gluon sector
in this talk), there are four of these distributions, 
$H$, $E$, $\tilde H$ and $\tilde E$,
which depend on three variables: the average longitudinal momentum
fraction $x$ of the struck parton in the initial and final states,
a skewness parameter $\xi$ which measures the difference between these
two momentum fractions, and the momentum transfer $t$ to the target nucleon.
The above mentioned factorization theorems~\cite{Ji97,Col97}
are applicable for $|t|/Q^2\ll 1$.

The physical content of the GPDs is quite rich:
in simple terms, whereas ordinary parton distributions yield,
for example, the probability
$|\psi(x)|^2$ that a quark carries a fraction $x$ of the
momentum of a fast moving nucleon, the GPDs measure the {\sl coherence} 
$\psi^*(x-\xi)\cdot\psi(x+\xi)$ between two different (quark) momentum states
of the nucleon and in this way quark momentum correlations in the
nucleon. 
In quantum mechanics, one cannot reduce the knowledge of a system
formed of a superposition of states (in our case, an infinite number of states
described by the variable $x$) to the sole probabilities that the
system be in a particular state. The interferences between these states,
or {\sl coherences}, or off-diagonal elements, are necessary for a full
description of the system. 

If one of the momentum fractions ($x+\xi$ or $x-\xi$) is negative,
it is interpreted as representative of an antiquark. 
Meson-like $q\bar{q}$ configurations
in the nucleon may then be investigated. As mentioned in Sec.~\ref{sec:models},
the GPDs are expected to exhibit a rich structure at $|x|<\xi$ which is
related in particular to the pion cloud around the nucleon. 

Whereas $x$ and $\xi$ characterize
solely the longitudinal momenta of the partons involved, the $t$-dependence
of the GPDs is related to their transverse momenta. By Fourier transform,
it is conceivable to access simultaneously the longitudinal momentum
fraction of quarks and their position in the transverse plane~\cite{Bur00}. This would
open the way to a {\sl femto-photography} of the nucleon~\cite{Ral01}.

The richness of information contained in the GPDs may be illustrated by
several quite remarkable relations, among which
\begin{itemize}
\item {\sl the forward limit:}
\[
\lim_{t\rightarrow 0,\xi\rightarrow 0}\, H(x,\xi,t)=q(x),
\ \hbox{or} \ -\bar{q}(-x)\ \hbox{if}\ x<0
\]
\centerline{and}
\[\lim_{t\rightarrow 0,\xi\rightarrow 0}\tilde{\, H}(x,\xi,t)=\Delta q(x),
\ \hbox{or} \ \Delta \bar{q}(-x).
\]
As expected from the above discussion, the ordinary parton distributions,
both unpolarized $q(x)$ and polarized $\Delta q(x)$,
are but a limiting case of the GPDs. 
Note that, though the GPDs are defined functions for $\xi=0$ or $t=0$, these
variables take only finite, non-zero values in any experiment. 
Note also that the functions $E$ and \( \tilde{E} \) have no connection
with the ordinary parton distributions. They are not constrained
by deeply inelastic scattering, which corresponds to this forward limit.

\item {\sl the form factor limit:}
\[\forall \xi,\ \ \ 
\int_{-1}^{+1} H(x,\xi,t)\cdot dx=F(t)  \]
 (4 relations of this type for each quark flavor). 
After integration over \( x \), the $t$-dependence
of the GPDs is given by the nucleon elastic form factors ($H\to$ Dirac, 
$E\to$ Pauli, $\tilde H\to$ axial vector and $\tilde E\to$ pseudoscalar
form factors).

\item {\sl Ji's sum rule:}
\[\forall \xi,\ \ \
\lim_{t\rightarrow 0}\frac{1}{2}\int_{-1}^{+1} x\cdot (H+E)(x,\xi,t)\cdot dx = J_{q}, 
 \]
where $J_{q}$ is the total angular momentum, 
i.e. the sum of intrinsic spin and of orbital angular momentum,
carried by the quarks. 
This sum rule provides a way to understand the
origin of the nucleon spin. Indeed, the
contribution of the quark intrinsic spin (\( \frac{1}{2}\Delta \Sigma  \))
was measured at CERN/SMC, SLAC and DESY/\-HER\-MES. The gluon contribution ($\Delta G$)
will be determined at CERN/COMPASS, RHIC/STAR and SLAC. 
The knowledge of GPDs
allows one to isolate the contribution of the quark orbital angular momentum
to the total spin of the nucleon: 
$\frac{1}{2} = J_q + J_g=(\frac{1}{2}\Delta \Sigma + L_q)+(\Delta G+L_g)$.

\end{itemize}

The first two relations demonstrate the new unifying frame given by the
GPDs: observables as different as elastic form factors and parton distributions
are related to each other through these new functions. However, the knowledge of
limits, or projections, does not imply a complete knowledge of the
GPDs themselves. Models are being developed for a complete representation.
\section{GPD calculations}
We give here a short summary of selected calculations of the GPDs. For a
more complete account, see Refs.~\cite{Goe01,Bel02,Vdh02}.

\subsection{Constrained parametrizations}
\label{sec:parametrizations}
The ``double distributions"~\cite{Rad99} are well
suited for the description of the GPDs. They incorporate the GPDs
mathematical properties (polynomiality of the Mellin moments) and
have a definite physical content. 
They satisfy
positivity bounds, a number of inequalities that GPDs
have been shown to obey. Elastic form factors and parton distributions
are included as limiting values.

An alternate, ``dual" parametrization, based on a partial wave expansion,
has recently been proposed~\cite{Pol02}.

\begin{table*}
\caption{Comparative merits of DVCS and DVMP for the study of Generalized Parton
	Distributions}
\label{tab:DES}
\begin{center}
\begin{tabular}{|c|c|}
\hline 
\textbf{  DVCS}{  }&
\textbf{  DVMP}{  }\\
\hline 
\hline
{  Factorization established }&
{  Factorization established for longitudinal photons}\\
{  (virtual photon is transverse at leading order)}&
{  (\( \sigma_{L} \) must be extracted experimentally)}\\
\hline 
{  }&
{  Vector mesons : combination of \( H \) and \( E \)}\\
{  }&
{  (the GPD's entering Ji's sum rule)}\\
{  Measures a combination of \( H,E,\tilde{H},\tilde{E} \)}&
{  Pseudoscalar mesons : combination of \( \tilde{H} \) et
\( \tilde{E} \)}\\
{  }&
{  Various decompositions in quark flavours }\\
{  }&
{  according to the nature of the meson}\\
\hline 

{  \( \sigma_T(\gamma^*p\to\gamma p) \propto Q^{-4} \)}&
{  \( \sigma_L(\gamma^*p\to p\ \hbox{meson})\propto Q^{-6} \)}\\
&
{  (larger at least for \( \pi ^{+} \), \( \rho  \) and \( \omega  \)
		at moderate \( Q^{2} \))}\\
\hline 
{  Contribution of Bethe-Heitler process:}&
{  }\\
{  - useful through interferences}&
\\
{  (polarization or charge asymmetries measurements)}&
\\
{  - decreases quickly with increasing energy}&
{  }\\
\hline
{  }&
{  A manageable complication: the meson structure}\\
\hline  
{  Smaller higher twist contributions}&
{  Larger higher twist contributions, but cancel}\\
&
{  in ratios \( \rho /\omega  \) and polarization asymmetries}\\
\hline  

\end{tabular}
\end{center}
\end{table*}

\subsection{Model calculations}
\label{sec:models}
QCD-inspired models are very useful to understand how non-perturbative
mechanisms generate various structures in GPDs. 

In the simple bag model,
GPDs have but a weak dependence on $\xi$, and show no structure for $|x|<\xi$.
A number of calculations in different constituent quark models are 
being developed.

In contradistinction, in the chiral quark-soliton model, where the
quarks interact through a highly non-linear pion field, the GPDs exhibit
a rich structure for $|x|<\xi$. The same model leads also to a
$t$-dependence which can be approximated by $H(x,0,t)\sim x^{-at}q(x)$
and deviates significantly from the early factorized {\sl ansatz}
$H(x,0,t)\sim F(t)q(x)$. 

In the region $x>\xi$, the use of light-cone wave functions for the
nucleon allows to write GPDs as overlap integrals. In considering
the nucleon as a superposition of Fock states, e.g. 
$|p> = |uud> + |uudq\overline{q}> +|uudg> + \cdots$, the GPDs are
useful to understand the role of given states.

\subsection{Lattice QCD}
\label{sec:LQCD}
Moments of ordinary parton distributions are
being evaluated using QCD calculations on a lattice. 
Likewise, GPD moments can be calculated~\cite{Sch02}.

\section{Observables}
\label{sec:obs}
Both DVCS and DVMP processes are sensitive to GPDs of the nucleon, once it
is established that the handbag diagram of Fig.~\ref{fig:handbag} dominates
in the kinematical regions where measurements are performed. 
These processes
have different merits and sensitivities summarized in Table~\ref{tab:DES}.
The sensitivity of specific observables, including polarization observables,
has been studied in detail in Refs.~\cite{Goe01,Bel02}. 

While $\xi$ and $t$ are variables fixed by the kinematics of the process
under study, $x$ enters the amplitude as an integration variable. Therefore,
in most cases, observables are expressed in terms of integrals over $x$ of
GPDs. The only exception is in observables sensitive only to the imaginary part
of the handbag amplitude: because of the propagators $(x\pm\xi+i\varepsilon)^{-1}$
in the diagram of Fig.~\ref{fig:handbag}, the GPDs enter then with an argument
$x=\pm\xi$. This is the case for the DVCS beam spin asymmetries discussed in
Sec.~\ref{sec:exp}. 

In general, the extraction of GPDs from observables
requires a deconvolution and, in principle, an infinite set of data. In practice,
this problem has mathematical properties analogous to  e.g. tomography~\cite{Ter01}
and may be solved (with associated uncertainties) with a finite set of data. 
This complete exercise remains to be done with either real or simulated data.

\section{Finite $Q^2$}
\label{sec:q2}
Most of the discussion so far was based on a high momentum transfer  
approximation (Bjorken limit). In practice, the measurements of
the exclusive processes under study are or will be performed at
rather moderate $Q^2$, 1 to 6 GeV$^2$. The relevant corrections are listed
in this Section.

As is the case for the ordinary parton distributions, the GPD evolution
leads to logarithmic corrections ($\ln Q^2$): the dependence
of the GPDs on a factorization scale $\mu$ results from the evolution 
equation \hfill \\
$\ \ \ \mu \frac{\partial}{\partial\mu}H(x,\xi,t;\mu)=\int {\cal K}(x,y,\xi;\alpha_s(\mu))
	\cdot H(y,\xi,t;\mu) \cdot dy$,\\ 
where $\alpha_s$ is the strong coupling constant.	
The kernel ${\cal K}$ has been calculated at next-to-leading order (NLO).

At order ${\cal O}(1/Q)$ (twist-3), beyond a minimal gauge restoring term, the
contribution of a longitudinal ex\-chan\-ged photon $\gamma^*_L$ to DVCS can be expressed
in terms of derivatives of (twist-2) GPDs. This is the so-called Wand\-zu\-ra-Wilczek 
approximation. The exchange of a gluon between the quark in the handbag loop and
the nucleon core would give an additional contribution, thought to be very small,
in analogy with a similar case in deeply inelastic scattering.

At the next order in $Q$, trivial kinematical
corrections ${\cal O}(t/Q^2,M^2/Q^2)$, surprisingly enough, are not
always considered. There are also dynamical target mass corrections ${\cal O}(M^2/Q^2)$.
Finally the transverse momenta of quarks modify the quark propagators in the diagram
of Fig.~\ref{fig:handbag}: 
$(x\pm\xi+i\varepsilon)^{-1} \to (x\pm\xi+k_{\perp}^2/Q^2+i\varepsilon)^{-1}$.
Many other twist-4 contributions remain to be evaluated.

The aim of the first generation experiments (see Sec.~\ref{sec:exp}) is then
to establish that observables have the anticipated $Q^2$ dependence in order to
validate the GPD approach at leading twist or to disentangle higher twist
contributions.

\section{Experiments}
\label{sec:exp}

The difficulty in the experimental study of GPDs is the measurement of exclusive
processes, of rather low cross sections, at the highest possible momentum transfer
$Q^2$. This requires high beam energy and luminosity, as well as large acceptance
and high resolution detectors. At present, these studies have been initiated at
HERMES (sufficient energy, 27 GeV, but relatively low luminosity) and at 
Jefferson Lab (beam energy of 4 to 6 GeV, but high product luminosity $\times$ acceptance).
In the next few years, significant progress could be achieved with a
12 GeV upgrade at JLab/CEBAF~\cite{12gev}, and using the 200 GeV muon beam at 
COMPASS~\cite{dHo00}.
A new high luminosity, high duty cycle machine, operating in the 25-50 GeV
range, would be ideally suited for these studies~\cite{Fer02}.

DVCS should be the cleanest process for GPD studies. Through the optical theorem,
it is linked to deeply inelastic scattering for which the evolution of the
structure functions is well understood down to $Q^2\sim 1$ GeV$^2$. 
In addition, the
$\gamma\gamma^*\pi^0$ form factor, whose calculation proceeds through the
same type of handbag diagram as in Fig.~\ref{fig:handbag}, gives also a
good indication that DVCS should scale early. Corrections may then be
expected to be understood at rather low $Q^2$. As mentioned in Table~\ref{tab:DES},
DVCS is indistinguishable from the known Bethe-Heitler (BH) process where the final 
photon is emitted by the incident or scattered electron. DVCS dominates over BH
only at very high energies (COMPASS, HERA). At beam energies accessible at
JLab or HERMES, the large BH process may be turned into an advantage by measuring
interferences in spin or charge asymmetries; DVCS could then be accessed at the
amplitude level.

\begin{figure}
\vspace{2mm}
\centerline{\epsfig{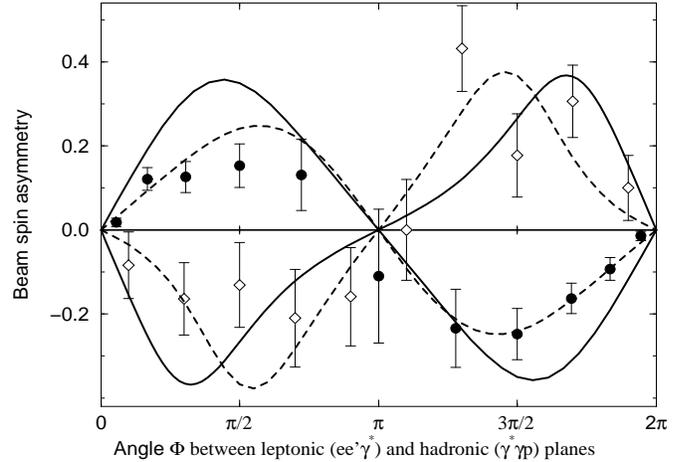}}
\vspace{2mm}
\caption{ 
	 Beam spin asymmetry in $\vec{e}p\to ep\gamma$.
	 The diamonds are results from HERMES~\cite{Air01}, 
	 the full circles from JLab/CLAS~\cite{Ste01}.	 
	 The kinematics are different for the two experiments. 
	 The corresponding calculations are 
	 from Refs.~\cite{Goe01} (solid lines) and~\cite{Bel02} (dashed lines). 
	 The asymmetry is of opposite sign between the two experiments
	 because one uses a positron beam while the other uses electrons. 
	 The approximate $\sin\Phi$ dependence is strongly suggestive of 
	 the validity of the GPD approach at leading order.
	}
\label{fig:clas+hermes}
\end{figure}

The first experimental indications of DVCS at the quark level were published
last year~\cite{Air01,Ste01}. DVCS was also observed at HERA~\cite{hera}, 
in a kinematical region that makes it more sensitive to gluons. The approximate
$\sin\Phi$ dependence of the observed beam spin asymmetry 
(see Fig.~\ref{fig:clas+hermes}) is expected from
any process which conserves helicity in the $\gamma^*p\to \gamma p$ 
scattering~\cite{Die97}.
The process described by the (leading order) handbag diagram of Fig.~\ref{fig:handbag}
has such a property. The  comparison with first 
calculations~\cite{Goe01,Bel02,Fre02} using the GPD formalism
is very encouraging. From the experimental point of view, it should be noted that
at HERMES, the 800 MeV missing mass resolution in the $ep\to e\gamma X$ reaction
is not sufficient to select unambiguously the (dominating) $ep\to ep\gamma$ events,
while at JLab/CLAS, the 110 MeV missing mass resolution in the $ep\to ep X$ reaction
causes the $ep\to ep\gamma$ and $ep\to ep\pi^0$ to mix, but a separation was done
using a shape analysis of the relevant spectra. 
First results of beam charge asymmetry measurement in DVCS were also reported
from HERMES~\cite{vdS02}.

Both experiments will soon
obtain more statistics.
Furthermore, both detection systems will be improved
in the next couple years. In particular, at JLab/CLAS, a forward 
lead-tungstanate calorimeter, covering the photon angles between 3 and 12$^{\circ}$,
is being built, together with a specially designed two-coil solenoid acting
as a trap for background M\"oller electrons~\cite{Bur01}. A luminosity
of $2\times 10^{34}$ cm$^{-2}$s$^{-1}$ is anticipated with this
large acceptance spectrometer. 
Another
DVCS experiment will run in 2003 at JLab/Hall A~\cite{Ber00}, using one High-Resolution
Spectrometer for detection of the scattered electrons, a new lead-fluoride 
calorimeter for the outgoing photons, and a ring of thick plastic scintillators 
for the recoil protons~\cite{Ber00}. 
Preliminary tests indicate that a luminosity of 
$10^{37}$ cm$^{-2}$s$^{-1}$ can be reached, in spite of the rather
open geometry and forward disposition of the proton detector.

Because of the necessary additional gluon exchange to create a meson, the
asymptotic regime should be reached at higher $Q^2$ in DVMP processes.
However observables such as asymmetries and ratios (e.g. $\rho/\omega$)
are expected to scale earlier.
As illustrated in Table~\ref{tab:DES}, DVMP, at leading order, is sensitive
to different combinations of GPDs as DVCS. Also the different quark content
of various mesons gives access to different flavour combinations of these
distributions. Since factorization has been proven only for longitudinal
photons~\cite{Col97}, the corresponding contribution must be extracted from the
experiment. For vector me\-sons, this is done via a measurement of the
meson polarization through the angular distribution of its decay products
($\rho\to\pi^+\pi^-$, $\omega\to\pi^+\pi^-\pi^0$, $\phi\to K^+K^-$) and
the verifiable application of $s$-channel helicity conservation to relate
the virtual photon helicity to the one of the emitted vector meson. For
pseudoscalar mesons, one has to resort to a Rosenbluth separation to extract
$\sigma_L$ (this has not been attempted yet in the considered kinematical
regime).

The first results of (almost) exclusive electroproduction of $\rho$ mesons 
(longitudinal cross sections)~\cite{Air00}
in a kinematical region where the handbag diagram is expected to dominate are
also very suggestive of the validity of the GPD approach. Systematic studies
and tests of the predicted $Q^{-6}$ scaling
of the longitudinal cross sections
are underway at JLab/CLAS: data taken at beam energies of 4.2 GeV~\cite{Had02} 
and 5.75 GeV~\cite{Gui99}, up to $Q^2\sim 4$ GeV$^2$, are being analyzed.

\begin{figure}
\vspace{-17mm}
\centerline{\epsfig{file=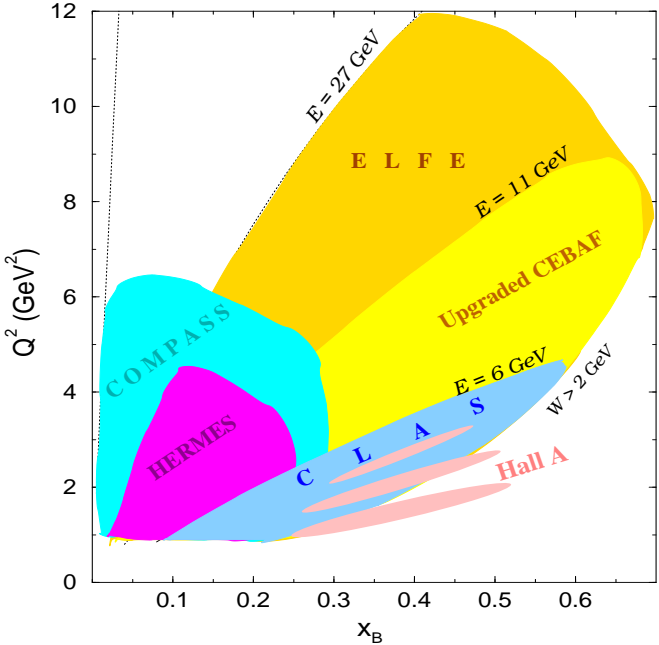,height=120mm,width=101mm}}
\vspace{-16mm}
\caption{
	 Kinematical domain ($Q^2$ vs $x_B$) accessible at various facilities
	 (existing or in project) for the measurement of deeply virtual exclusive
	 processes. $x_B$ is the usual Bjorken variable, from which $\xi$ is
	 calculated. ``ELFE" does not refer here to a specific project, but
	 to a - yet to be defined - high luminosity machine, in this case at 27 GeV.
	 See text for additional comments.
	}
\label{fig:q2_xb}
\end{figure}

The kinematical domain accessible with existing and projected facilities is
illustrated in Fig.~\ref{fig:q2_xb}. The high $Q^2$ approximate limit for the 
existing high energy facilities results from count rate limitations 
due to a lower luminosity and decreasing cross sections as $Q^2$ and $x_B$ increase.

\section{Conclusions and outlook}

Generalized Parton Distributions have emerged in the past few years as a powerful,
attractive and unifying concept for the hadron structure. Their $Q^2$ evolution
has been worked out to next-to-leading order, and twist-3 (and some twist-4)
contributions to deeply virtual and exclusive processes have been calculated.
The interpretation and significance of GPDs are under intense study.

The deconvolution problem posed by the extraction of GPDs from experimental observables
was barely touched upon, but mathematical methods exist, or are being developed,
that address this particular issue. 

On the experimental side, we are still at a very early stage. There are indications
that the handbag diagram is at work. In the next two or three years, more dedicated
and higher statistics experiments will attempt to establish on firm grounds the validity
of this theoretical approach; this includes tests of $Q^2$-scaling laws and of factorization.
Beyond that, systematic measurements of GPDs will be performed in a large kinematical
domain at several facilities.


\begin{thebibliography}{}

\bibitem{Vdh99} M. Vanderhaeghen, P.A.M. Guichon and M. Guidal,
		Phys. Rev. D {\bf 60} (1999) 094017;
		P.A.M. Guichon and M. Vanderhaeghen,
		Prog. Part. Nucl. Phys. {\bf 41} (1998) 125-190.
		
\bibitem{Fil01} B.W. Filippone and Xiandong Ji,
		Adv. Nucl. Phys. {\bf 26} (2001) 1-88.
		
\bibitem{Goe01} K. Goeke, M.V. Polyakov and M. Vanderhaeghen,
		Prog. Part. Nucl. Phys. {\bf 47} (2001) 401-515.
		
\bibitem{Rad01} A.V. Radyushkin, hep-ph/0101225.		
		
\bibitem{Bel02} A.V. Belitsky, D. M\"uller and A. Kirchner,
		Nucl. Phys. {\bf B629} (2002) 323-392.

\bibitem{Vdh02} M. Vanderhaeghen, these proceedings.

\bibitem{Ji97}	X. Ji, Phys. Rev. D {\bf 55} (1997) 7114-7125;
		A.V. Radyushkin, Phys. Rev. D {\bf 56} (1997) 5524-5557.

\bibitem{Col97}	J.C. Collins, L. Frankfurt, and M. Strikman, 
		Phys. Rev. D {\bf 56} (1997) 2982-3006.

\bibitem{Bur00} M. Burkardt,
		Phys. Rev. D {\bf 62} (2000) 071503(R).

\bibitem{Ral01} J.P. Ralston and B. Pire, hep-ph/0110075;
		M. Diehl, Eur. Phys. Jour. C {\bf 25} (2002) 223-232;
		J.P. Ralston and P. Jain, hep-ph/0207129;
		M. Burkardt, hep-ph/0209179.
		
\bibitem{Rad99} A.V. Radyushkin,
		Phys. Rev. D {\bf 59} (1999) 014030.
		
\bibitem{Pol02} M.V. Polyakov and A.G. Shuvaev, hep-ph/0207153.

\bibitem{Sch02} W. Schroers, 
		Workshop on Hadronic phenomenology from lattice gauge theory,
		Univ. Regensburg, August 1-3, 2002.


\bibitem{Ter01} O.V. Teryaev, Phys. Lett. B {\bf 510} (2001) 125-132.

\bibitem{12gev} The Science Driving the 12 GeV Upgrade of CEBAF,
		Jefferson Lab report, February 2001.
		
\bibitem{dHo00} N. d'Hose {\sl et al.}, Letter of intent to the CSTS/SPhN-Saclay (2000),
		http://www-dapnia.cea.fr/Sphn/Vcs/Deep/.
		
\bibitem{Fer02} {\sl Declaration of Ferrara}, 
		http://www.fe.infn.it/qcd-n02/.

\bibitem{Air01} A. Airapetian {\sl et al.} (HERMES collaboration), 
		Phys. Rev. Lett. {\bf 87} (2001) 182001.
		
\bibitem{Ste01} S. Stepanyan {\sl et al.} (CLAS collaboration),
		Phys. Rev. Lett. {\bf 87} (2001) 182002.

\bibitem{hera}  C. Adloff {\sl et al.} (H1 collaboration),
		Phys. Lett. {\bf B517} (2001) 47-58.
		
\bibitem{Fre02} A. Freund, M. McDermott and M. Strikman,
		hep-ph/0208160.
		
\bibitem{Die97} M. Diehl {\sl et al.},
		Phys. Lett. {\bf B411} (1997) 193-202.
		
\bibitem{vdS02} G. van der Steenhoven, these proceedings.

\bibitem{Bur01} V. Burkert, L. Elouadrhiri, M. Gar\c con, S. Stepanyan {\sl et al.},
		CEBAF experiment 01-113.

\bibitem{Ber00} P. Bertin, C. Hyde-Wright, R. Ransome, F. Sabati\'e {\sl et al.},
		CEBAF experiment 00-110.

\bibitem{Air00}	A. Airapetian {\sl et al.} (HERMES collaboration), 
		Eur. Phys. Jour. C {\bf 17} (2000) 389-398.
		
\bibitem{Had02} C. Hadjidakis, private communication.
		
\bibitem{Gui99} M. Guidal, M. Gar\c con, E. Smith {\sl et al.},
		CEBAF experiment 99-105.
		
\end{thebibliography}
\end{document}